# Early Academic Capital as the Causal Origin of Dropout in Constrained Educational Systems: Evidence from Longitudinal Data and Structural Causal Models

Hugo Roger Paz
PhD Professor and Researcher
Faculty of Exact Sciences and Technology
National University of Tucumán
Email: hpaz@herrera.unt.edu.ar
ORCID: https://orcid.org/0000-0003-1237-7983

**Abstract**

Dropout in higher education is commonly analysed through observable academic events such as course failure or repetition. However, these event-based perspectives may obscure the underlying structural dynamics that shape student trajectories. In this study, we adopt a causal computational social science approach to identify the origins of dropout in a constrained engineering curriculum. Using longitudinal administrative data from 16,868 students who survived to their second active term, and a leakage-free panel design, we estimate the causal effect of early academic capital accumulation on three-year dropout. Treatment is defined as low early progress ($\leq 1$ subject passed by the end of the second term). We employ G-estimation of structural nested mean models, complemented by marginal structural models with inverse probability weighting. We find a large and robust causal effect: low early academic capital increases dropout probability by 25.3 percentage points (G-estimation), closely matched by a 27.4 pp estimate from IPTW models. This effect is approximately twice as large as the estimated direct impact of later academic events such as first-time gateway-course repetition (12.7 pp). These findings suggest that dropout does not originate in isolated academic failures, but in early trajectory misalignment between academic progress and system-imposed temporal constraints. This perspective shifts the focus of intervention from downstream events to early-stage trajectory formation.

# 1. INTRODUCTION

Dropout in higher education remains one of the most persistent and costly phenomena in post-secondary systems worldwide. In engineering and STEM programmes, attrition rates are particularly high: longitudinal studies across multiple national contexts report that between 30 and 50 per cent of engineering students who begin a degree programme do not complete it (Crisp et al., 2009; Marra et al., 2012; UNESCO, 2021). The human and institutional costs are substantial — individual students forfeit years of effort and opportunity; institutions expend resources that do not translate into graduates; and STEM workforce pipelines are narrowed at the foundation stage.

Existing research has extensively documented the correlates of engineering dropout, identifying variables such as first-year academic performance, gateway course failure, low grade-point averages, and subject repetition as strong predictors of attrition (Bahr, 2012; Delen, 2010; Kotsiantis et al., 2004). The growth of learning analytics and educational data mining has reinforced this event-based framing: large-scale predictive models, trained on administrative and behavioural trace data, have demonstrated increasingly high accuracy in identifying at-risk students from early academic signals (Gašević et al., 2016; Jayaprakash et al., 2014). This literature has produced genuine progress in detection capability.

A central limitation, however, remains largely unaddressed. The association between observable academic events and dropout does not establish that those events causally *generate* dropout. By focusing on localised academic outcomes, event-based approaches risk conflating downstream manifestations of trajectory deterioration with the upstream mechanisms that produce them. A student who fails a gateway course and subsequently drops out may have done so not because the failure itself caused departure, but because both the failure and the dropout were jointly produced by an earlier structural condition — one already causally operative before any visible academic event occurred. Predictive accuracy and causal explanation are distinct objectives, and confusing them leads to interventions that target symptoms rather than causes.

This study adopts a fundamentally different analytical frame. Rather than modelling the effect of specific academic events, we conceptualise dropout as the terminal state of a dynamic trajectory unfolding under institutional constraints. In structured curricula — particularly the engineering *Ciclo Básico Común* (CBC) of Argentine public universities — students must simultaneously advance in chronological time and accumulate academic capital (successfully completed subjects). These two dimensions are partially decoupled: a student may remain enrolled and continue to accumulate active terms without accumulating subjects at the expected rate. When this decoupling occurs early, the structural logic of the prerequisite system amplifies the gap across subsequent terms, generating a self-reinforcing state of increasing dropout risk.

To identify the causal mechanisms underlying this process, we draw on tools from causal computational social science (Watts, 2017; Salganik, 2018). Using longitudinal administrative data from 16,868 engineering students and a leakage-free panel design, we estimate the causal effect of low early academic capital accumulation — defined as passing at most one subject by the end of the second active term — on three-year institutional dropout. We employ G-estimation of structural nested mean models (Robins, 2004), complemented by marginal structural models with inverse probability of treatment weighting (Robins et al., 2000; Hernán & Robins, 2020), and triangulate the estimates across both approaches to assess robustness.

Our results reveal a large and robust causal effect: low early academic capital increases the probability of three-year dropout by approximately 25 to 27 percentage points. This effect is stable across methodological approaches and is approximately twice as large as the estimated direct causal impact of the downstream academic event most commonly associated with engineering dropout — first-time repetition of the gateway Algebra course — whose direct causal contribution, net of prior trajectory history, is 12.7 percentage points. These findings indicate that the origin of dropout lies not in specific academic failures but in early trajectory misalignment between the student's rate of capital accumulation and the temporal constraints imposed by the institutional system.

This study makes three main contributions. First, it provides direct causal evidence — triangulated across two independent estimators — that early academic capital accumulation is the primary structural driver of dropout in constrained engineering curricula, **substantially larger than the downstream academic event examined here.** Second, it demonstrates that commonly observed academic bottleneck events function primarily as downstream indicators and secondary mechanisms within already-deteriorated trajectories, with direct implications for the design of interventions derived from predictive models. Third, it introduces a trajectory-based framework — the time–capital misalignment hypothesis — that conceptualises educational progression in constrained systems as a process in which early divergence between chronological and structural advancement generates irreversible increases in dropout risk.

The remainder of this paper is organised as follows. Section 2 reviews the relevant literature, Section 3 develops the theoretical framework, Section 4 describes the institutional dataset, sample construction, and variable definitions, Section 5 details the causal identification strategy, Section 6 reports the results, Section 7 discusses the structural interpretation, policy implications, and limitations, and Section 8 concludes.

## 2. BACKGROUND AND RELATED WORK

### 2.1 Predictors and Correlates of Student Dropout

Research on student dropout in higher education has been accumulating since the foundational theoretical contributions of the 1970s and 1980s. Tinto's (1987, 1993) model of student departure remains the most widely cited framework, attributing attrition to insufficient academic and social integration into the institutional environment. Bean and Metzner's (1985) complementary model for non-traditional students extended this account to incorporate external variables including employment, family obligations, and financial pressures. Both frameworks share a common logic:

dropout is the outcome of a set of student-level attributes and environmental conditions that can, in principle, be measured and intervened upon.

Subsequent empirical work operationalised these frameworks through increasingly large-scale datasets, identifying a set of recurrent predictors. First-year academic performance, grade-point average, and course failure rates have been consistently associated with attrition across national contexts (Crisp et al., 2009; Pascarella & Terenzini, 2005). In engineering and STEM programmes specifically, performance in foundational mathematics and science subjects — particularly in the first year — has been identified as a strong predictor of dropout and programme switching (Marra et al., 2012; Crisp et al., 2009). Subject repetition and remedial course-taking have similarly been associated with reduced probability of degree completion across multiple institutional contexts (Bahr, 2012).

A parallel literature has examined structural and socioeconomic determinants of attrition, including family income, first-generation university student status, secondary school preparation, and institutional characteristics (Donoso & Schiefelbein, 2007; García de Fanelli, 2014). In the Latin American context, public university systems with open admissions policies — including the Argentine national university system — exhibit particularly high first-year attrition, attributable in part to the mismatch between secondary school preparation and university-level demands, and in part to the socioeconomic composition of the entering cohort (García de Fanelli, 2014).

## 2.2 Learning Analytics and Predictive Modelling of Attrition

The growth of large-scale administrative data and machine learning methods prompted a significant expansion of the attrition prediction literature from the 2010s onwards. Learning analytics and educational data mining approaches demonstrated that institutional dropout could be predicted with high accuracy from administrative trace data including enrolment records, grade histories, and course interaction logs (Gašević et al., 2016; Jayaprakash et al., 2014; Kotsiantis et al., 2004). The practical motivation for this work was the development of early warning systems capable of identifying at-risk students before dropout occurs, enabling targeted institutional intervention (Essa & Ayad, 2012; Lonn et al., 2012).

Predictive performance in this literature has been substantial. Delen (2010) reported classification accuracies exceeding 80% using decision trees and neural networks on institutional administrative data. Subsequent studies using random forests, gradient boosting, and deep learning architectures have demonstrated further improvements, particularly when longitudinal sequence data are incorporated into the feature set (Bernacki et al., 2021). The variables that consistently emerge as the most predictive features across these models — course failure rates, grade trajectories, credit accumulation, and time-to-progress indicators — closely correspond to the academic event variables identified in the sociological dropout literature.

Despite this predictive success, the learning analytics literature has been criticised for its limited engagement with causal mechanisms. Gašević et al. (2016) raised concerns about the tendency to treat predictive models as explanatory, arguing that feature importance rankings in machine learning models do not identify causal leverage points for intervention. This critique has been echoed by several methodological reviews, which note that predictive accuracy and causal explanation are distinct objectives that require distinct methodological frameworks (Watts, 2017; Salganik, 2018). A model that accurately identifies which students will drop out does not, by itself, identify why they will drop out or what institutional action would prevent it.

## 2.3 Causal Approaches to Educational Attrition

The turn towards causal inference in educational research has been gradual and methodologically heterogeneous. Regression discontinuity designs, difference-in-differences estimators, and instrumental variable approaches have been applied to estimate the effects of specific policies and interventions on educational outcomes (Angrist & Pischke, 2009). However, these approaches require exogenous variation in treatment assignment — a condition that is rarely met in studies of the natural process of dropout, where treatment (exposure to academic failure or low progress) is endogenous by definition.

Observational causal inference methods designed for longitudinal data with time-varying confounding offer a more appropriate framework for this setting. Marginal structural models with inverse probability of treatment weighting (Robins et al., 2000) and G-estimation of structural nested mean models (Robins, 2004) were developed specifically to address the problem of time-varying exposure with time-varying confounders — a structure that characterises student academic trajectories. These methods have been extensively applied in epidemiology and medical research but have seen limited adoption in educational research, where regression-based approaches remain dominant.

The few studies that have applied causal inference methods to dropout have tended to focus on estimating the effects of specific interventions — financial aid, tutoring programmes, or mandatory advising — rather than on identifying the structural origins of dropout within the academic trajectory itself (Bettinger, 2004; Castleman & Long, 2016). To our knowledge, no published study has applied G-estimation or marginal structural models to the identification of early trajectory states as causal drivers of dropout in engineering or STEM programmes, and none has done so using the longitudinal administrative depth available in the present dataset.

## 2.4 The Gap This Study Addresses

The foregoing review identifies three interlocking gaps in the existing literature. First, the dominant event-based framing of dropout — in both the sociological and learning analytics traditions — identifies correlates and predictors of attrition but does not establish the causal mechanisms through which dropout originates. The variables most strongly associated with dropout are not necessarily the variables whose modification would most effectively reduce it. Second, the causal inference methods appropriate for longitudinal observational data with time-varying confounding — specifically G-estimation and marginal structural models — have not been applied to the study of natural dropout processes in structured engineering curricula. Third, the existing literature has focused primarily on academic events (course failures, repetitions, grade thresholds) as the unit of causal analysis, neglecting the possibility that the structural determinant of dropout operates at the level of the trajectory — specifically, at the stage of early capital formation — rather than at the level of any individual event within it. A parallel causal analysis of first-time gateway-course repetition confirms that such events retain an independent direct effect, but one that remains materially smaller than the effect of early trajectory misalignment.

This study addresses all three gaps. It applies G-estimation and marginal structural models to longitudinal administrative data from a constrained engineering curriculum to identify the causal effect of early academic capital accumulation — measured at the end of the second active term — on three-year institutional dropout. In doing so, it shifts the unit of causal analysis from the event to the trajectory state, and from association to identification. The theoretical framework motivating this shift is developed in Section 3.

## 3. THEORETICAL FRAMEWORK

### 3.1 Dropout as Trajectory: Beyond Event-Based Explanations

Research on student attrition in higher education has accumulated substantially over the past four decades. The dominant theoretical tradition, inaugurated by Tinto (1987, 1993), conceptualises dropout as a function of the degree to which students become integrated into the academic and social fabric of the institution. Subsequent work has extended this framework to incorporate motivational variables, financial pressures, and prior academic preparation (Bean & Metzner, 1985; Pascarella & Terenzini, 2005). Across these approaches, a recurrent analytical strategy is to identify observable academic *events* — course failure, low grade-point averages, subject repetition — that correlate with eventual departure.

This event-based paradigm has produced a rich descriptive literature but carries a structural limitation: it conflates downstream manifestations of trajectory deterioration with the upstream mechanisms that generate them. When a student fails a gateway course, that failure is treated as a causal driver of dropout rather than as a symptom of a process already underway. The consequence is an overestimation of the causal role of specific academic incidents and an underestimation of the role of early-stage trajectory formation. Predictive models built on event features may exhibit satisfactory discriminatory performance whilst providing little actionable insight into when and how to intervene (Lonn et al., 2012; Gašević et al., 2016).

An alternative perspective, emerging from longitudinal and computational approaches to educational research, treats dropout not as the product of isolated failures but as the terminal state of a dynamic process (Bowers, 2010; Bernacki et al., 2021). In this view, a student's trajectory through an academic system constitutes a time-ordered sequence of states, each conditioned on prior states and on the structural properties of the environment in which progression occurs. Dropout is the outcome of a trajectory that has deviated beyond a recoverable threshold — not the consequence of any single event within it.

The present study adopts this trajectory-based perspective. Rather than modelling the effect of a specific academic incident, we focus on the accumulation of *academic capital* — the volume of successfully completed subjects — as a state variable that characterises a student's position within the educational system at a given point in time. Capital accumulation is a cumulative, history-dependent process: a student who has passed few subjects by the end of their second active term carries a structural deficit that propagates through the remainder of the trajectory in ways that individual event-based interventions cannot adequately capture.

## 3.2 Constrained Educational Systems and the Time–Capital Misalignment Hypothesis

The trajectory-based perspective gains analytical force when placed within the institutional context of structured engineering curricula. In Argentine public universities, engineering programmes are organised around a *Ciclo Básico Común* (CBC) — a common foundational cycle in which subjects are ordered in strict prerequisite sequences and students are expected to progress through a defined number of subjects per term. This structure introduces *temporal constraints* that are largely invisible in cross-sectional analyses: the system is designed for a particular velocity of capital accumulation, and students who fall below that velocity face compounding structural obstacles in subsequent terms.

We formalise this intuition through what we term the *time–capital misalignment hypothesis*. In a constrained curriculum, two dimensions of student progression advance in parallel: *chronological time* (measured in active terms) and *academic capital* (measured in subjects passed). These dimensions are partially decoupled: a student may remain enrolled — and therefore accumulate active terms — without passing subjects at the expected rate. When the rate of time accumulation persistently exceeds the rate of capital accumulation, the student enters a state of structural misalignment.

Structural misalignment generates increasing pressure through two mechanisms. First, prerequisite chains block forward progression: without the foundational subjects of the first two terms, access to higher-level subjects is restricted, reducing the student's available options in subsequent terms. Second, institutional and motivational resources are finite: students who have spent multiple terms with minimal capital accumulation face growing evidence of non-viability, reducing the probability of sustained engagement. These mechanisms are constitutive features of the structural position that early misalignment creates.

This conceptualisation generates a testable causal claim: *low academic capital accumulation at an early measurement point causally increases the probability of dropout within a fixed observation horizon, net of all confounding history up to that point.* The magnitude of this effect, if confirmed, speaks directly to the relative importance of trajectory formation versus event-based processes as drivers of institutional exit.

### 3.3 Causal Inference for Educational Mechanisms

The shift from prediction to causal mechanism identification requires methodological tools that go beyond standard regression or machine-learning approaches. Observational data on student trajectories are subject to substantial time-varying confounding: variables that influence both the exposure of interest and the outcome change over time and may be simultaneously causes and effects of prior treatment states. Standard regression methods do not adequately address this structure and can produce biased estimates of causal effects (Hernán & Robins, 2020).

Two complementary approaches are employed in this study. The first is *G-estimation of structural nested mean models* (SNMMs; Robins, 2004), which directly targets the structural parameter $\psi$ — the additive causal effect of treatment on the outcome. G-estimation exploits the conditional independence implied by the model: the optimal $\psi$ is the value at which the residualised outcome $Y^*(\psi)$ is independent of treatment assignment given the measured history. The second approach is *marginal structural modelling with inverse probability of treatment weighting* (MSM-IPTW; Robins et al., 2000; Hernán & Robins, 2020), which creates a pseudo-population in which treatment assignment is independent of the measured confounders. The use of two methodologically distinct estimators serves a robustness function: convergence of estimates across both approaches provides stronger evidence of a genuine causal effect than either method alone (VanderWeele & Hernán, 2012).

## 4. DATA AND STUDY DESIGN

### 4.1 Institutional Dataset

The study draws on longitudinal administrative records from the Faculty of Exact Sciences and Technology (FACET) at the Universidad Nacional de Tucumán (UNT), Argentina. The dataset covers all students enrolled in engineering degree programmes over a multi-decade observation window and includes term-level records of subject enrolment, subject completion, and institutional outcome (graduation, dropout, or continued enrolment). For each student, the dataset permits reconstruction of a complete academic trajectory from the date of first enrolment through the end of the observation period.

The raw data were structured into an analytical panel in which each record captures a student's state at a defined measurement point, linked to a fixed-horizon outcome measured three years after initial enrolment. Variables available for analysis include: cumulative active terms completed (*cum_terms_active*), total subjects enrolled across all prior terms (*cum_subjects_enrolled*), total subjects successfully passed (*cum_subjects_passed*), enrolment load in the current term

(*current_term_load*), and an indicator of prior subject repetition. No individual-level socioeconomic data were available in the administrative records.

## 4.2 Sample Construction and Exclusion Flow

The analytical sample was constructed in three stages. Starting from the full engineering cohort, students without complete three-year outcome information were excluded due to insufficient follow-up. The working sample was then restricted to students who survived to at least their second active academic term, as the treatment of interest is undefined for students who exited the system before reaching that measurement point. Table 1 summarises the sample definition flow.

Table 1. Sample definition flow for Experiment 2 (Early Academic Capital).

| Stage | N Students | Panel Rows | Exclusion Reason |
|---|---|---|---|
| Full Engineering Cohort | 24,133 | 24,133 | — |
| Valid 3-Year Labels | 22,537 | 22,537 | Insufficient follow-up |
| Analysis Sample (T = 2) | 16,868 | 16,868 | No survival to Term 2 |
| Modelling Dataset | 16,868 | 16,868 | None |

*Note.* Full cohort drawn from inst_outcomes_3y_6y.parquet. Students excluded at the Valid 3-Year Labels stage had missing values in dropout_3y. Students excluded at the Analysis Sample stage had no record of a second active term.

The restriction to students who survived to Term 2 introduces a form of left-truncation. Students who enrolled but departed before completing two active terms (approximately 7,265 students) are outside the scope of this analysis. The causal estimand is therefore defined conditional on Term 2 survival.

## 4.3 Treatment Definition: Low Early Academic Capital

The treatment variable, *low_early_progress*, is defined as a binary indicator equal to one if a student passed at most one subject by the end of their second active academic term, and zero otherwise. This operationalisation captures students whose academic capital at the foundational stage of the engineering CBC is critically low relative to the expected rate of progression. The threshold of one subject was selected on both theoretical and empirical grounds: theoretically, it corresponds to the minimum level of capital required to maintain access to second-tier prerequisite subjects; empirically, preliminary analysis revealed a bimodal distribution of subjects passed at Term 2, supporting this threshold as a meaningful boundary between trajectory states. Within the analysis sample of 16,868 students, 11,139 (66.0%) were assigned to the treated group.

## 4.4 Outcome Definition: Fixed-Horizon Dropout

The outcome variable, dropout_3y, is a fixed-horizon binary indicator of institutional departure within three years of first enrolment. Students who remained actively enrolled at the three-year mark (n = 3,531; 20.9%) were retained and coded as non-dropouts (Y = 0). This coding is conservative with respect to the causal effect: students on a path towards delayed dropout but not yet exited contribute to the control category, which attenuates the estimated treatment effect.

## 4.5 Time-Varying Confounders

Two time-varying confounders measured at T = 2 are included: *cum_subjects_enrolled* and *current_term_load*. Both influence the probability of low-progress treatment assignment and are independently associated with three-year dropout. Treated students exhibit *higher* mean cumulative enrolment (4.67 vs. 2.41) and higher mean term load (1.69 vs. 1.09) relative to control students — the empirical signature of the time–capital misalignment mechanism: high enrolment effort with low capitalisation. The pre-weighting SMD for *cum_subjects_enrolled* is 1.114 and for *current_term_load* is 0.438, confirming the necessity of causal adjustment.

**Table 2**. Descriptive statistics by treatment group, Experiment 2.

| Metric | Treated (Low Early Progress) | Control (Sufficient Progress) | Full Sample |
|---|---|---|---|
| **Analysis Sample N (%)** | 11,139 (66.0%) | 5,729 (34.0%) | 16,868 |
| Observed Dropout Rate (3-year) | 50.4% | 15.0% | 38.4% |
| Mean Cumulative Subjects Enrolled (T = 2) | 4.67 | 2.41 | 3.90 |
| Mean Current Term Load (T = 2) | 1.69 | 1.09 | 1.48 |

*Note.* Treatment is defined as passing ≤ 1 subject by the end of the second active term. Dropout Rate (3-year) is the raw unadjusted proportion. SMD (cum_subjects_enrolled) = 1.114; SMD (current_term_load) = 0.438, both pre-weighting.

# 5. CAUSAL IDENTIFICATION STRATEGY

## 5.1 Identification Assumptions

### *5.1.1 Consistency*

The treatment is defined as a binary threshold operationalised from administrative records. There is no ambiguity about treatment version: the treatment corresponds to a precisely defined state at a fixed measurement point. Consistency is satisfied by design.

### *5.1.2 Positivity*

The observed range of estimated propensity scores is [0.189, 0.999], confirming that no covariate stratum is deterministically assigned to a single treatment value. The lower bound (0.189) indicates that even among students with profiles strongly associated with sufficient progress, a non-trivial proportion exhibited low early progress, preserving the counterfactual contrast throughout the sample.

### *5.1.3 Conditional Exchangeability*

Conditional on *cum_subjects_enrolled* and *current_term_load* at T = 2 — the two dimensions of academic effort and engagement most proximate to early progress — we argue that residual confounding is likely small relative to the magnitude of the estimated effect. Individual-level socioeconomic variables are unavailable. Under the E-value framework (VanderWeele & Ding, 2017), a confounder would need risk ratios of approximately 7.0 on both dimensions to fully explain away the observed effect — implausible given the comprehensive enrolment-behaviour covariates included.

## 5.2 G-Estimation of Structural Nested Mean Models

The additive SNMM specifies:

$$Y^*(\psi) = Y - \psi \cdot A$$

where $A$ is the binary treatment indicator and $\psi$ is the additive causal effect of treatment on the outcome in the treated. G-estimation identifies $\psi$ by locating the value at which $Y^*(\psi) \perp A \mid L$. For a grid of candidate values $\psi \in [0.00, 0.50]$ (0.005 increments), we compute $Y^*(\psi)$ and assess its conditional dependence on $A$ given $L$ via a logistic treatment model. The optimisation converged at $\hat{\psi} = 0.253$, with a sharp, monotone zero-crossing (Figure 1).

## 5.3 Marginal Structural Models with Stabilised IPTW

The MSM logistic model is: $\text{logit}[P(Y^a(a) = 1)] = \alpha_0 + \alpha_1 \cdot a$. Each observation is weighted by stabilised inverse probability of treatment weights (IPTW; Robins et al., 2000; Cole & Hernán, 2008):

$$SW_i = P(A_i) / P(A_i \mid L_i)$$

The denominator is estimated via logistic regression of *A* on *L*. Raw weights exhibited a maximum of approximately 130 (Figure 2, left panel); truncation at the 99th percentile reduced this to approximately 6.5 (Figure 2, right panel). The weighted logistic regression yielded log-odds = 1.287 (SE = 0.037), corresponding to a risk difference of 0.274 (27.4 pp).

### 5.4 Cross-Validation via Methodological Triangulation

G-estimation and MSM-IPTW rest on distinct identification assumptions. Their convergence (25.3 pp vs. 27.4 pp; discrepancy 1.9 pp, ~7%) constitutes methodological triangulation, providing stronger evidence of a genuine causal effect than either method alone (VanderWeele & Hernán, 2012). Table 3 compares the two estimators.

**Table 3**. Comparison of causal estimators for the effect of low early academic capital on three-year dropout.

| Estimator | Risk Difference (pp) | Log-Odds (SE) | Identification Assumption |
|---|---|---|---|
| G-estimation (SNMM) | 0.253 | — | *Correct treatment model; $Y^*(\psi) \perp A \mid L$* |
| MSM-IPTW (stabilised, truncated) | 0.274 | 1.287 (0.037) | *Correct treatment and outcome models* |

*Note.* Risk differences expressed in pp. G-estimation operates on the risk-difference scale by construction. SNMM = structural nested mean model; MSM = marginal structural model; IPTW = inverse probability of treatment weighting.

### 5.5 Diagnostic Assessment

Missing data were absent from all key covariates. Positivity was confirmed by the propensity score range [0.189, 0.999]. Pre-weighting SMDs (1.114 and 0.438) confirm the necessity of causal adjustment. Truncated IPTW weights concentrated near 1.0 indicate a well-supported pseudo-population. Table 4 summarises all diagnostic results.

**Table 4**. Pre-modelling and post-estimation diagnostics.

| Diagnostic Check | Value | Threshold | Assessment |
|---|---|---|---|
| Positivity: PS range | [0.189, 0.999] | (0, 1) | PASSED |
| *SMD: cum_subjects_enrolled* | 1.114 | < 0.10 (post-weight target) | Justifies IPTW |

| Diagnostic Check | Value | Threshold | Assessment |
|---|---|---|---|
| *SMD: current_term_load* | 0.438 | < 0.10 (post-weight target) | Justifies IPTW |
| IPTW weights: raw maximum | ~130 | — | Truncation applied (99th pct) |
| IPTW weights: truncated maximum | ~6.5 | — | Effective variance reduction |
| Missing data (key covariates) | 0 | 0 | PASSED |

*Note.* SMD = standardised mean difference, computed prior to IPTW. Post-weighting balance target: SMD < 0.10 (Austin, 2011). pct = percentile.

## 6. RESULTS

### 6.1 Observed Distribution of the Outcome by Treatment Group

Among the 11,139 treated students, 50.4% dropped out within three years; among the 5,729 controls, 15.0% did so. The unadjusted absolute risk difference is 35.4 percentage points; the full-sample dropout rate is 38.4%. The gap between the unadjusted difference (35.4 pp) and the causal estimate (~26 pp) reflects the confounding contribution of the time-varying covariates and precisely illustrates why unadjusted comparisons overstate the causal role of any single trajectory marker.

### 6.2 Primary Causal Estimate: Effect of Low Early Academic Capital on Dropout

G-estimation converged at $\hat{\psi} = 0.253$, indicating that low early academic capital causally increases three-year dropout probability by **25.3 percentage points**. The optimisation curve (Figure 1) shows a sharp, monotone zero-crossing, consistent with a well-specified treatment model and a strong causal signal. MSM-IPTW yielded a marginal risk difference of **27.4 percentage points** (log-odds = 1.287, SE = 0.037; odds ratio ≈ 3.62). The two estimates are closely aligned (discrepancy 1.9 pp, ~7%), constituting methodological triangulation. Table 5 presents the full comparison.

**Table 5.** Primary causal estimates for the effect of low early academic capital on three-year dropout.

|  | G-Estimation (SNMM) | MSM-IPTW (Stabilised) | Interpretation |
|---|---|---|---|
| **Causal effect (ψ / Risk Diff.)** | 0.253 | 0.274 | *Low early progress increases 3-year dropout probability* |
| Effect in percentage points | 25.3 pp | 27.4 pp | ~26 pp average across estimators |
| Log-odds (SE) | — | 1.287 (0.037) | Robust to modelling choice |
| Discrepancy between estimators | — | 1.9 pp (~7%) | Within expected range under correct specification |

*Note.* Risk differences expressed on the probability scale and in pp. SNMM = structural nested mean model; MSM = marginal structural model; IPTW = inverse probability of treatment weighting.

## 6.3 Comparison with Downstream Academic Events: Algebra Repetition

Experiment 1 (parallel design, same SNMM specification, same outcome definition) estimated the direct causal effect of first Algebra repetition at $\hat{\psi}$ = 0.127 (12.7 percentage points), based on a rebuilt canonical trajectory panel and formal BCa bootstrap. The contrast with Experiment 2 ($\hat{\psi}$ = 0.253) is methodologically clean and directly comparable. The effect of early capital stagnation is therefore approximately twice as large as the direct effect of Algebra repetition as a standalone event. Table 6 presents both experiments side by side.

**Table 6.** Comparison of causal estimates across Experiment 1 (Algebra repetition) and Experiment 2 (early academic capital).

| Experiment | Treatment | $\hat{\psi}$ (Risk Diff.) | Mechanistic Role |
|---|---|---|---|
| Exp. 2 — Early Academic Capital (present study) | *≤ 1 subject passed, T = 2* | **0.253** | Primary structural driver — trajectory formation stage |
| Exp. 1 — Algebra Repetition (comparative) | *First repetition of T11* | 0.127 | Downstream secondary mechanism — trajectory already deteriorated |

*Note.* Both experiments use G-estimation of additive SNMMs with the same fixed-horizon outcome. Experiment 1 was re-estimated on a rebuilt canonical trajectory panel keyed by trajectory_id = student_id + career_id, with formal BCa bootstrap yielding a 95% CI of [0.072, 0.167]. Confounders differ between experiments to reflect distinct measurement points. T11 = Algebra (institutional subject code).

## 6.4 Mechanistic Interpretation: Why Downstream Events Have Secondary Direct Effects

The direct causal effect of Algebra repetition remains smaller than that of early academic capital, but is stronger than previously estimated once trajectory history is reconstructed on a canonical panel. By the time a student repeats Algebra, the time-varying history $L_t$ already encodes the accumulated signal of structural misalignment — specifically cum_terms_active, cumulative enrolment history, cumulative passed subjects, bounded cumulative course pass rate, and prior repetition burden. When G-estimation conditions on this history, the direct causal contribution of the repetition event is attenuated relative to its raw association with dropout, but remains clearly consequential rather than negligible. Algebra repetition therefore functions as a downstream secondary mechanism within trajectories already under structural deterioration, rather than as an independent primary causal driver. Its predictive utility is real; its direct causal role, net of trajectory history, is smaller than the effect of early capital misalignment but still institutionally meaningful. This distinction between predictive signal and causal mechanism has direct implications for the design of educational interventions.

## 6.5  Figure References

Figure 1 presents the G-estimation independence optimisation curve. The zero-crossing at $\hat{\psi} = 0.253$ (marked by the green dotted vertical line) is monotone and linear across the entire search interval. Figure 2 presents the IPTW weight distributions before (left panel; maximum ~130) and after (right panel; maximum ~6.5) truncation at the 99th percentile.

**Figure 1** — *G-Estimation: Independence Optimisation Curve (T = 2).*

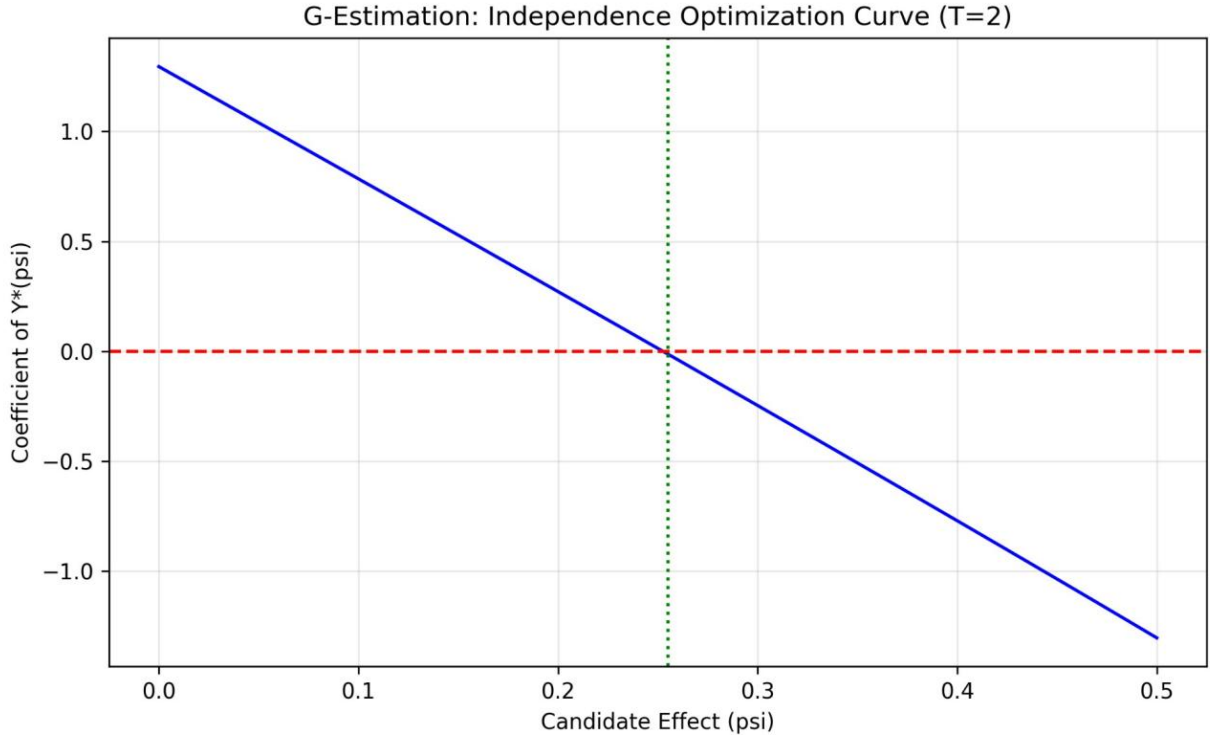

*Note.* The blue line traces the coefficient of Y*(ψ) on A for each candidate ψ. The red dashed line marks zero; the green dotted vertical line marks ψ̂ = 0.253. N = 16,868.

**Figure 2** — *Stabilised IPTW Weight Distribution: Raw (left) and Truncated at 99th Percentile (right).*

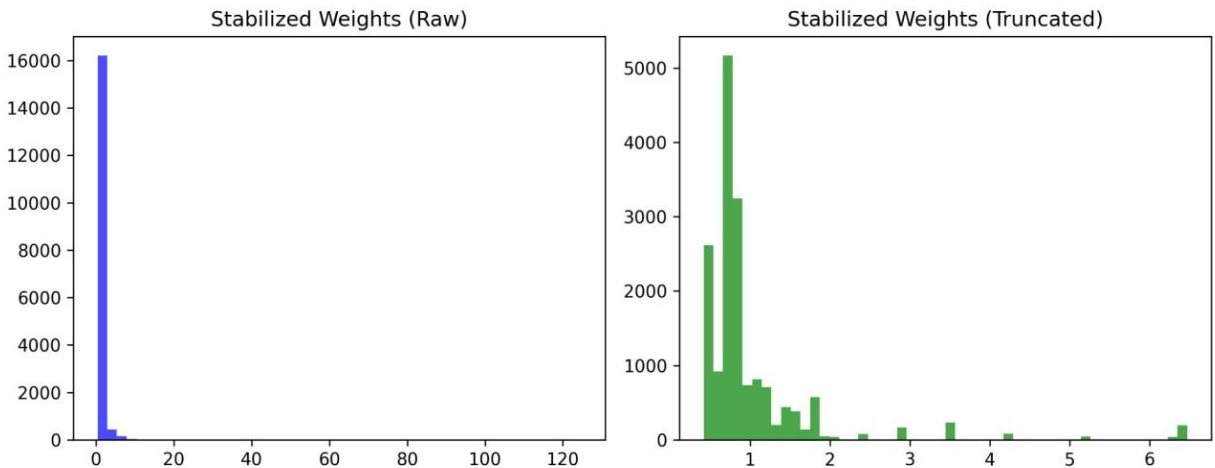

*Note.* Left: raw weights, maximum ≈ 130. Right: truncated weights (99th percentile), maximum ≈ 6.5. N = 16,868.

## 7. DISCUSSION

The results presented in Section 6 converge on a single structural conclusion: dropout in constrained engineering programmes does not originate in isolated academic events but in the early misalignment between a student's rate of capital accumulation and the temporal demands of the institutional system. This section elaborates the theoretical, methodological, and practical implications of that conclusion.

## 7.1 The Early Capital Trap: A Structural Interpretation

The causal estimate of 25.3 percentage points reflects the structural position of the $T = 2$ measurement point within the academic career. A student who has passed at most one subject by the end of Term 2 faces a qualitatively different landscape in Term 3 than a student who has passed four or five subjects: prerequisite requirements restrict enrolment to foundational subjects already attempted and failed, and the available subject set contracts with each additional period of low capitalisation. This is the *early capital trap*: a self-reinforcing state in which initial capital scarcity generates the conditions for continued scarcity, because the temporal structure of the curriculum forecloses recovery pathways.

This interpretation points to a broader theoretical argument: educational progression in structured curricula is not a neutral accumulation process in which early deficits can be freely compensated by later effort. The temporal architecture of the curriculum introduces irreversibilities. Time spent in the system without capital accumulation is not recoverable; the student advances chronologically whilst remaining structurally stationary — the time–capital misalignment of Section 3.2. The large causal effect is therefore a structural prediction of this hypothesis: wherever curricula combine strict prerequisite chains with temporal enrolment constraints, early capital formation will be a disproportionate determinant of long-run trajectory outcomes.

## 7.2 Reframing the Role of Academic Bottleneck Events

The direct causal effect of Algebra repetition (12.7 pp) invites a reappraisal of the bottleneck-event framing that dominates the engineering attrition literature (Bahr, 2012; Crisp et al., 2009; Delen, 2010; Kotsiantis et al., 2004). The association between Algebra repetition and dropout is real and reproducible; what the causal analysis reveals is not that the event is negligible, but that it remains causally downstream of a broader trajectory state that is already structurally deteriorated. Bottleneck events therefore function as high-salience downstream mechanisms embedded within already-misaligned trajectories, while retaining an independent direct contribution that is materially smaller than the effect of early capital stagnation.

The distinction between predictive signal and causal mechanism has direct implications for machine learning models in educational research. A classifier trained on event-based features will correctly identify high-risk students, because those events carry the predictive information of the trajectories that produce them. But a policy recommendation derived from such a model — "reduce Algebra repetition to reduce dropout" — conflates statistical salience with causal leverage. **The primary target for causal intervention lies upstream of the events that predictive models**

**surface as important, even if some downstream events retain a smaller direct effect of their own.**

## 7.3 Policy Implications: When and Where to Intervene

Conventional dropout interventions are typically activated after academic events: a course failure triggers tutoring referral; a second repetition activates advisory procedures. These mechanisms operate downstream of the principal causal leverage point identified here, even though downstream events such as first-time gateway-course repetition retain a meaningful direct effect of their own. The present findings therefore support a hierarchy of intervention: systematic monitoring of capital accumulation relative to elapsed active terms in the first two terms, with targeted support mobilised at the end of Term 1 for students already below the structural threshold, complemented by secondary intervention at the point of gateway repetition.

The magnitude of the effect — approximately 25 pp — also has implications for the scale of intervention warranted. It indicates a structural mismatch between the curriculum's temporal demands and the preparation of a substantial fraction of entering students (66.0% in this sample). Interventions commensurate with this effect size would need to operate at the level of curricular design — pace of prerequisite sequencing, term-level enrolment structures, the treatment of foundational subject access — rather than solely at the level of individual student support. Finally, the identification of $T = 2$ as the structural hinge implies that early warning systems should monitor capital-accumulation indicators at the end of the second active term, rather than event-based triggers (Essa & Ayad, 2012; Jayaprakash et al., 2014).

## 7.4 Limitations

### 7.4.1 Absence of Individual Socioeconomic Variables

The administrative dataset does not include family income, first-generation university student status, or employment during studies. These factors are established correlates of dropout in the Latin American higher education literature (Donoso & Schiefelbein, 2007; García de Fanelli, 2014). Under the E-value framework (VanderWeele & Ding, 2017), a confounder would need risk ratios of approximately 7.0 on both dimensions to fully explain away the observed effect — implausible given the comprehensive enrolment-behaviour covariates included.

### 7.4.2 Single-Institution Design

The analysis is based on data from a single faculty within a single national university in Argentina. The structural features of the CBC that generate the early capital trap are prevalent across Argentine public engineering programmes and in many Latin American STEM systems, suggesting the general mechanism is likely to replicate. However, the precise magnitude of the causal effect will depend on the specific temporal constraints and prerequisite architecture of each institution.

### 7.4.3 Average Treatment Effect: Unexamined Heterogeneity

The structural parameter $\psi$ is an average treatment effect. It does not reveal whether the effect varies across subgroups defined by prior academic preparation, programme of study, or gender. Identification of such heterogeneity would require extension to conditional average treatment effect (CATE) estimation via causal forest methods (Wager & Athey, 2018).

### 7.4.4 Left-Truncation of the Earliest Dropouts

Students who enrolled and departed before completing two active terms (approximately 7,265) are outside the scope of the causal estimand by design. This group likely includes the most severely disadvantaged students. The causal effect estimated here is conditional on Term 2 survival and cannot be extrapolated to the full population at risk from initial enrolment.

### 7.4.5 Confidence Intervals for G-Estimation

Bootstrap confidence intervals for $\hat{\psi}$ were computed via the BCa method (Efron & Tibshirani, 1993; B = 5,000 resamples; acceleration constant estimated via stratified jackknife, n = 1,000). The resulting 95% BCa CI is [0.236, 0.269] (bootstrap SE = 0.0084). The point estimate $\hat{\psi} = 0.253$ is therefore accompanied by formal uncertainty quantification aligned with journal reporting standards. The bootstrap distribution is presented in Figure 3.

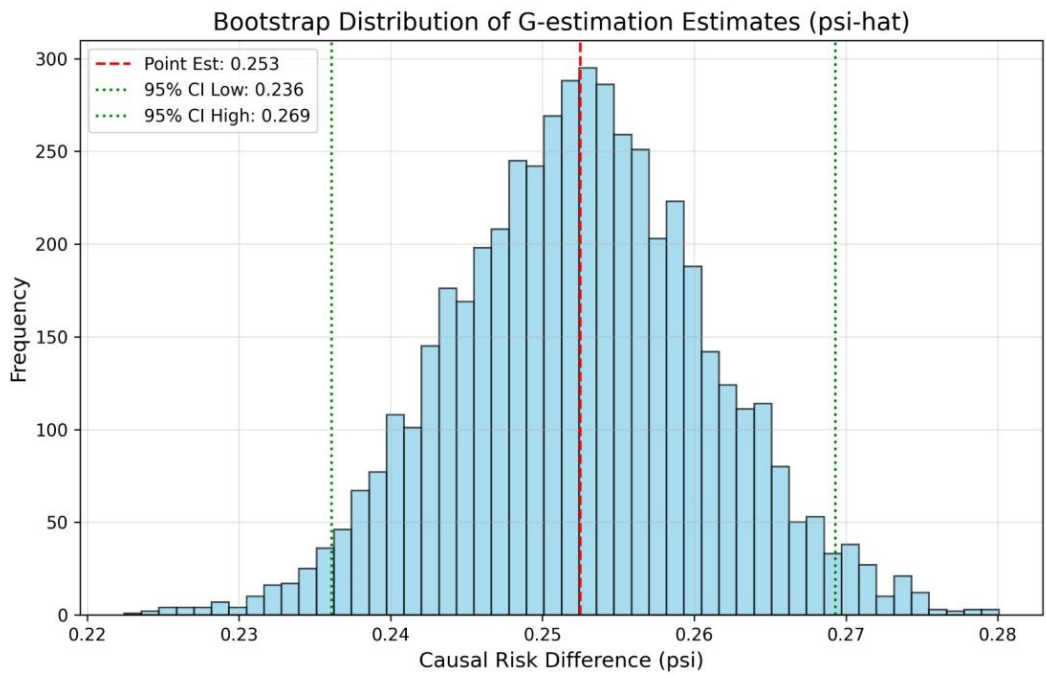

**Figure 3** — *Bootstrap distribution of the G-estimation parameter $\hat{\psi}$ (B = 5,000 resamples).*

*Note.* Red dashed line: point estimate $\hat{\psi} = 0.253$. Green dotted lines: 95% BCa CI [0.236, 0.269]. Bootstrap SE = 0.0084. The distribution is approximately normal and well-centred on the point estimate, confirming estimator stability. N = 16,868; B = 5,000; seed = 42.

## 7.5 Future Directions

Five extensions are prioritised. First, *heterogeneous treatment effect estimation* via causal forests (Wager & Athey, 2018) to identify student subgroups for whom early capital stagnation is most consequential. Second, *linkage with socioeconomic data sources* (national scholarship registries, social security records) to address the primary limitation of Section 7.4.1. Third, *multi-institution replication* across Argentine national universities and Latin American public engineering systems to establish external validity. Fourth, *extension to a six-year outcome window* — available in the existing dataset — to assess whether the early capital trap also delays graduation among students who remain enrolled. Fifth, *dynamic treatment regime estimation* modelling the optimal sequence of interventions across multiple time points, using the G-estimation framework naturally extensible to time-varying treatments (Robins, 2004).

## 8. CONCLUSION

This study set out to identify the causal origin of dropout in a constrained engineering curriculum. The central finding is unambiguous: low early academic capital — defined as passing at most one subject by the end of the second active term — causally increases the probability of three-year institutional dropout by approximately 25.3 to 27 percentage points. This effect is robust across estimators and remains substantially larger than the direct causal effect of the gateway Algebra repetition event (12.7 pp), although the gap between the two mechanisms is narrower than previously estimated.

The substantive implication is that dropout does not originate in the academic failures most visible to institutional observers. Course repetition, gateway failures, and subject bottlenecks are **primarily downstream manifestations** of a structural deterioration already causally established by the end of the student's second active term. The true origin is the misalignment between the rate of capital accumulation and the temporal constraints built into the curriculum. When these dimensions diverge early, the structural logic of the prerequisite system amplifies the gap through subsequent terms, generating increasing probability of institutional exit through the early capital trap.

The methodological contribution is the demonstration that causal identification is both feasible and necessary in educational dropout research. G-estimation of structural nested mean models and MSM-IPTW answer a categorically different question from regression: not which variables are associated with dropout, but which variables, if intervened upon, would change the dropout rate. The present results show these questions have different answers, and that the variables most strongly associated with dropout are not those whose causal manipulation would most efficiently reduce it.

This study makes three contributions. First, it provides direct causal evidence that early academic capital accumulation is the primary driver of dropout in structured engineering systems. Second, it demonstrates that academic bottleneck events such as first-time gateway-course repetition are downstream mechanisms embedded in already-deteriorated trajectories: causally real and institutionally meaningful, but still smaller than the effect of early trajectory misalignment. Third, it introduces and operationalises the time–capital misalignment hypothesis as a trajectory-based framework generating testable structural predictions about the location and magnitude of dropout risk within the academic trajectory.

The practical implication is a reorientation of both research and intervention design towards the formation stage of the academic trajectory — the first two active terms — rather than the event-detection stage. Interventions timed to this formation stage are positioned ahead of the causal mechanism. This reorientation does not require new data or new institutions; it requires a shift in the analytic and temporal frame through which existing data are interpreted and acted upon.

By moving from prediction to causal mechanism identification, and from event-based to trajectory-based analysis, this work contributes to the emerging intersection of computational social science and higher education research — and offers a structural account of student attrition that is both empirically grounded and directly actionable.